\documentclass[12pt]{article}

\usepackage[dvips]{graphicx}
\usepackage{epsfig}
\usepackage{amsmath,amsfonts,amssymb,amsthm}
\usepackage{verbatim}
\usepackage{psfrag}
\usepackage{bm}
\usepackage{bbm}
\usepackage[square,comma,sort&compress,numbers]{natbib}
\usepackage{color}
\usepackage{slashed}
\usepackage{axodraw}

\usepackage{epsf,epsfig}
\usepackage{graphics}

\begin{document}
\thispagestyle{empty}

\begin{flushright}
{
\small
TTK-11-14\\
}
\end{flushright}

\vspace{0.4cm}
\begin{center}
\Large\bf\boldmath
Self Regulation of Infrared Correlations for Massless Scalar Fields during Inflation
\unboldmath
\end{center}

\vspace{0.4cm}

\begin{center}
{Bj\"orn~Garbrecht and Gerasimos Rigopoulos}\\
\vskip0.3cm
{\it Institut f\"ur Theoretische Teilchenphysik und Kosmologie,\\
RWTH Aachen University,\\
D--52056 Aachen, Germany}\\
\vskip.3cm
\end{center}

\begin{abstract}
Self-energies of a minimally coupled scalar field with quartic
and trilinear interactions are calculated in a de~Sitter background, using a position space propagator. For quartic interactions, we recover earlier results for the seagull diagram, namely that it contributes an effective mass for the scalar field at leading order in the infrared enhancement in a steady-state de~Sitter background. We further show that the sunset diagram also contributes to this effective mass and argue that these two contributions are sufficient in order to determine a self-consistent dynamical mass. In addition, trilinear interactions also induce a dynamical mass for the scalar field which we calculate. Since an interacting scalar field in de~Sitter acquires a dynamical mass through these loop corrections, the infrared divergences of the two-point correlator are naturally self-regulated.

\end{abstract}


\section{Introduction}

It is well known that in de~Sitter space, there is no propagator
for a massless minimally coupled scalar field, which is of
Hadamard form and that exhibits all
ten space-time symmetries~\cite{Allen:1985ux,Allen:1987tz,Tsamis:1993ub}
at the same time.
The origin of this problem can be understood when constructing
the propagator as an integral of free field modes. The integration
over momentum space then leads to an infrared (IR) divergence.
For this reason, one often refers to this issue as the IR
problem of de~Sitter space.
({\it Cf.} Ref.~\cite{Seery:2010kh}, where this and other matters
are reviewed.)
Under certain circumstances, this
may lead to problems for models of the inflationary stage of cosmic
expansion, where the space-time is approximately de~Sitter.
A plausible and pragmatic
solution for the propagator during inflation relies on the
assumption that inflation begins a finite number of e-folds $N_{\rm e}$
before it terminates. Then one may use a propagator that
breaks de~Sitter symmetry through an IR-enhanced term proportional to
the logarithm of the scale factor or, equivalently, to the
number of e-folds since the onset of inflation.
This growing IR-term can be identified with the accumulation of superhorizon
modes.

The circumstances when the solution with the growing, de~Sitter
breaking propagator are no longer satisfactory are when $N_{\rm e}$
grows so large that the IR enhanced term in the propagator leads to
a
breakdown of perturbation theory in loop diagrams. The IR
problem is therefore generic for models of inflation that assume
a large number of e-foldings or even an infinite number, such as
eternal inflation.

It is very plausible however that the IR problem is self-regulatory:
A growing IR term for a scalar field $\phi(x)$
leads to a growing, dynamical mass
$m_{\rm dyn}$ from the self-energy,
which in turn limits the IR-enhanced term.
To our knowledge, the viability of this idea has been demonstrated
for a $\phi^4$ theory using three different approaches:
stochastic inflation~\cite{Starobinsky:1994bd},
Schwinger-Dyson equations~\cite{Riotto:2008mv} and the
functional renormalisation group~\cite{Burgess:2009bs}.
All these works agree on
the result for the self-regulating, dynamical mass $m_{\rm dyn}$
that arises from the seagull diagram. A related
approach has been developed for Euclidean de~Sitter space, which
is compact and therefore the IR divergence can be isolated in
the discrete zero mode~\cite{Rajaraman:2010xd}. In the present work, we demonstrate that such an IR self-regularization does take place for scalar fields with trilinear and quartic interactions, and that for quartic interactions both the seagull and the sunset diagrams contribute to $m_{\rm dyn}$ and are sufficient to determine it.

The technical device we use to obtain our results is based on the Schwinger-Dyson equations in the Closed Time Path
(CTP) formalism in a de~Sitter background, which we introduce in Section~\ref{section:CTP}. While these concepts are well known,
we feel that their reiteration is necessary here in order to introduce the
notations and conventions that are used in what follows. The first application, discussed in
Section~\ref{section:one-loop}, is the seagull diagram
of $\phi^4$ theory for which we reproduce the existing
result~\cite{Starobinsky:1994bd,Riotto:2008mv,Burgess:2009bs}, namely that the diagram generates a dynamical mass $m_{\rm dyn}$ whose value we determine through a simple self-consistency relation arising from the Schwinger-Dyson equations. Technically, the present approach differs from the one of Ref.~\cite{Riotto:2008mv} in that we use an expansion of the scalar propagator in powers of
$m^2_{\rm dyn}/H^2$ that starts at order minus one, and
that has been developed in Ref.~\cite{Prokopec:2003tm}.
Using this propagator, we avoid appealing to stochastic arguments
and the Fokker-Planck equation
for the calculation of the IR-enhancement effects as it is
done in Ref.~\cite{Riotto:2008mv}. We note here, that the seagull correction is particularly
simple, because it involves just one vertex, such that
its effects are manifestly local and there are no memory integrals.

This simplification does not apply for trilinear interactions, discussed in Section~\ref{section:triscalar}, and for which
the self energy is non-local. However, we observe that the leading (negative order) contribution in $m_{\rm dyn}^2/H^2$ to the convolution
integral between self energy and propagator appearing in
the Schwinger-Dyson equations, is proportional to the retarded Green
function of the covariant d'Alembertian operator. Acting with
the d'Alembertian on the Schwinger-Dyson equations from the left
yields local equations. This requires however the use of the ansatz
of a free field with constant mass $m_{\rm dyn}$ for the scalar
field $\phi(x)$. The effectively local form is therefore only valid
in time-independent situations, when the IR contribution is not
evolving. Otherwise, the convolution integrals, which are
sometimes referred to as memory integrals, cannot be simplified by this
method. The main result of Section~\ref{section:triscalar} is
a self-consistency relation determining $m_{\rm dyn}$ for
$\phi^3$ theory.
We note that in order to determine the effective dynamical
mass of spin-$\frac12$ fermions in de~Sitter space, a similar
method as presented in Section~\ref{section:triscalar} has been
applied before in Ref.~\cite{Garbrecht:2006jm}.

In Section~\ref{section:sunset}, we return to a $\phi^4$ interaction and consider the 2-loop sunset diagram. Since two
of the internal lines in the retarded self energy can be IR enhanced, the methods discussed
for the trilinear interactions in the previous section can be applied here
straightforwardly. Because of the simultaneous IR enhancement
of two propagators, it turns out that the
suppression due to the additional vertex is compensated when
compared to the seagull diagram of section~\ref{section:one-loop}. Therefore, both corrections
need to be included in the self-consistency relation that
determines $m_{\rm dyn}$ for $\phi^4$ theory and this is
one of the main results of Section~\ref{section:sunset}.
Besides, we argue that diagrams at higher loop order only lead
to subdominant contributions to $m_{\rm dyn}$ at leading order
in the IR enhancement. Section~\ref{section:conclusions} is left for additional comments and conclusions.

Before closing the Introduction, we should note that the complete
expressions for the self-energies in $\phi^4$ theory are
given in Refs.~\cite{Onemli:2002hr,Onemli:2004mb,Brunier:2004sb}.
In the present work, we do not give such a
complete account of all terms to second order in the vertices
of the self-energy, but we only extract the leading contributions
in $m^2_{\rm dyn}/H^2$, which are of negative order and therefore
IR enhanced. Taking account of these corrections, we show that there
are self-consistent, de~Sitter invariant equations for linear perturbations
and for the propagators, that determine $m^2_{\rm dyn}$. In this
sense, Refs.~\cite{Onemli:2002hr,Onemli:2004mb,Brunier:2004sb} and the present work are complementary.

\section{Scalar Propagator in de Sitter Space and the CTP Formalism}
\label{section:CTP}

The models we consider are defined through the action
\begin{align}
S=\int d^4x\, {\cal L}\,,
\end{align}
where the Lagrangian is given by
\begin{align}
\label{Lagrangian}
{\cal L}=
\sqrt{-g}\left[
g^{\mu\nu}(\partial_\mu \phi)(\partial_\nu \phi)
-\frac 12 m^2 \phi^2
\right]
+{\cal L}_{\rm int}
+\delta{\cal L}\,.
\end{align}
The first term is the Lagrangian for a free, massive minimally
coupled scalar field $\phi(x)$. Throughout the present work,
we assume that $\phi$ is light, by which we imply the relation
$m\ll H$, where $H$ is the Hubble expansion rate. For
loop corrections, this will
lead to IR-enhanced terms $\sim H^2/m^2$ or $\sim H^2/m_{\rm dyn}^2$,
where
$m_{\rm dyn}$ is a dynamically generated mass that we
require
to satisfy $m_{\rm dyn}\ll H$ as well.
Interaction terms are encompassed within ${\cal L}_{\rm int}$.
These give rise to ultraviolet (UV) divergences, which are renormalised by local
counterterms that are contained in $\delta {\cal L}$.
In the present work, we are interested in the effects of
IR enhancements $\sim H^2/m^2$ or $\sim H^2/m_{\rm dyn}^2$.
Radiative corrections proportional to these enhancement factors
turn out to be UV finite at leading order. We therefore omit a discussion
of the UV divergences in the present context, leave the form of
$\delta{\cal L}$ unspecified, and refer to
Refs.~\cite{Prokopec:2003tm,Garbrecht:2006jm,Onemli:2002hr,Onemli:2004mb,Brunier:2004sb,Garbrecht:2006df}
for detailed account of the concepts and technicalities of renormalisation
in de~Sitter background.

We use conformal coordinates $x^\mu$ with the metric tensor
\begin{align}
\notag
g_{\mu\nu}=a^2\eta_{\mu\nu}=a^2{\rm diag}(1,-1,-1,-1)\,.
\end{align}
The conformal time is given by $\eta=x^0$, and for
de~Sitter space, the scale factor is $a=-\frac1{H\eta}$ and
$\eta\in[-\infty;0]$. By defining
$\Delta x^2(x;x^\prime)=x_\mu\eta^{\mu\nu}x^\prime_\nu$,
the de~Sitter invariant length function is
\begin{align}
y(x;x^\prime)=a(\eta)a(\eta^\prime)H^2\Delta x^2
=-4\sin^2\left(\frac12 H \ell (x;x^\prime)\right)\,,
\end{align}
where we have indicated its relation with the geodesic
distance $\ell (x;x^\prime)$.

The free-field equation of motion obtained from the
Lagrangian~(\ref{Lagrangian}) is
\begin{align}
&a^4(-\nabla_x^2-m^2)\phi(x)
\\\notag
=&
-a^2\frac{d^2}{d\eta^2}\phi(x)
-2 a\left(\frac{da}{d\eta}\right)\frac{d}{d\eta}\phi(x)
+a^2{\vec\nabla}^2\phi(x)-a^4m^2\phi(x)
=0\,,
\end{align}
and the equation
for the free scalar propagator ${\rm i}\Delta^{(0)}(x;x^\prime)$ is
\begin{equation}
\label{freepropagator}
a^4(-\nabla_x^2- m^2){\rm i}\Delta^{(0)fg}(x;x^\prime)
=fg\delta^{fg}{\rm i}\delta^4(x-x^\prime)\,,
\end{equation}
where $\nabla_{x\mu}$ is the covariant derivative with respect
to $x$.
We define $z=1+\frac y4$ and $\{f,g\}=\{+,-\}$ are the CTP indices. Because of the de~Sitter invariance, it
is possible to express $a^4\nabla_x^2$ as an operator
of $z$ and derivatives with respect to $z$ only (or, alternatively, $y$).
One finds that
\begin{align}
a^4 H^2\left[
z(1-z)\frac{d^2}{dz^2}
+4(\frac12-z)\frac{d}{dz}-\frac{m^2}{H^2}
\right]{\rm i}\Delta^{(0)fg}(x;x^\prime)
=fg\delta^{fg}{\rm i}\delta^4(x-x^\prime)\,.
\end{align}
The exact solution of this equation can be given in terms of
a hypergeometric function. For the present purpose,
we expand in $y$ and $m^2/H^2$
and find~\cite{Prokopec:2003tm}
\begin{align}
\label{Delta0}
{\rm i}\Delta^{(0)fg}(x,x^\prime)=\frac{H^2}{4\pi^2}
\left\{
-\frac1{y^{fg}}-\frac12\log(-y^{fg})+\frac{3H^2}{2m^2}-1+\log 2+O\left(\frac{m^2}{H^2}\right)
\right\}\,,
\end{align}
where $y^{fg}=a(\eta)a(\eta^\prime)H^2{\Delta x^{fg}}^2$.
The term $\sim H^2/m^2$ within the curly brackets is the IR-enhancement
factor.
For comparison, the corresponding expression for a massless field is \cite{Allen:1985ux, Prokopec:2003tm}
\begin{align}
\label{Delta:m=0}
{\rm i}\Delta^{fg}_{\rm m=0}(x,x^\prime)=\frac{H^2}{4\pi^2}
\left\{
-\frac1{y^{fg}}-\frac12\log(-y^{fg})+\frac{1}{2}\log(a(\eta)a(\eta')) -\frac{1}{4}+\log 2
\right\}\,.
\end{align}
The desired behaviour at the coincident point $x=x^\prime$
arises from the definitions
\begin{align}
{\Delta x^{++}}^2(x;x^\prime)
=&(|\eta-\eta^\prime|-{\rm i}\varepsilon)^2-|\mathbf x -\mathbf x^\prime|^2\,,\\
{\Delta x^{+-}}^2(x;x^\prime)
=&(\eta-\eta^\prime+{\rm i}\varepsilon)^2-|\mathbf x -\mathbf x^\prime|^2\,,\\
{\Delta x^{-+}}^2(x;x^\prime)
=&(\eta-\eta^\prime-{\rm i}\varepsilon)^2-|\mathbf x -\mathbf x^\prime|^2\,,\\
{\Delta x^{--}}^2(x;x^\prime)
=&(|\eta-\eta^\prime|+{\rm i}\varepsilon)^2-|\mathbf x -\mathbf x^\prime|^2\,.
\end{align}

For a two-point function $G(x;x^\prime)$ [in the present work,
this can either be a free propagator $\Delta^{(0)}(x;x^\prime)$,
a full propagator $\Delta(x;x^\prime)$
or the self-energy $\Pi(x;x^\prime)$],
we identify the Wightman-type functions
\begin{align}
G^<(x;x^\prime)=G^{+-}(x;x^\prime)\,,\qquad
G^>(x;x^\prime)=G^{-+}(x;x^\prime)\,,
\end{align}
and the time-ordered and anti-time ordered functions
\begin{align}
G^T(x;x^\prime)=G^{++}(x;x^\prime)\,,\qquad
G^{\bar T}(x;x^\prime)=G^{--}(x;x^\prime)\,.
\end{align}
Moreover, we note that these can be combined to obtain the
advanced and the retarded
two-point functions as
\begin{subequations}
\label{def:retav}
\begin{align}
\label{CTP:advanced}
G^A&=G^T-G^>=G^<-G^{\bar T}\,,
\\
\label{CTP:retarded}
G^R&=G^T-G^<=G^>-G^{\bar T}\,.
\end{align}
\end{subequations}
Applying these relations to the retarded and
advanced propagators $\Delta^{(0)R,A}$, it is interesting to note that
these do not exhibit the IR-enhancement and will depend on the mass
only at the order $m^2/H^2$, which is beyond the scope of the
approximations made within the present work.

The proper self-energy can be defined as
\begin{align}
\label{def:Pi:proper}
\Pi^{ab}(x;x^\prime)
={\rm i}\frac{\delta\Gamma_2[\Delta]}{\delta\Delta^{ba}(x^\prime;x)}\,,
\end{align}
where $\Gamma_2$ is $-{\rm i}$ times
the sum of all two particle-irreducible (2PI) vacuum diagrams with full propagators $\Delta$ as internal lines.
The Schwinger-Dyson equations are
\begin{align}
\label{Schwinger:Dyson}
a^4(-\nabla^2_x-m^2){\rm i}\Delta^{ab}(x;x^\prime)
+{\rm i}c\int d^4w\,{\rm i}\Pi^{ac}(x;w){\rm i}\Delta^{cb}(w;x^\prime)
=\delta^{ab}{\rm i}\delta^4(x-x^\prime)\,.
\end{align}
Note that these are generically non-linear, because $\Pi$ is a functional
of $\Delta$. These equations are formally exact and describe the full
real-time evolution of the quantum system, but due to their non-linearity
approximations are needed in practice. Within the present work, we
seek for the main corrections in terms of the IR enhancement factor
$H^2/m^2$ or $H^2/m^2_{\rm dyn}$. Useful insight into the
dynamics may also be gained form an effective equation of motion for
the one-point function
\begin{align}
\label{EOM:effective}
a^4(-\nabla_x^2-m^2)\phi(x)-\int d^4 x^\prime\, \Pi^R(x;x^\prime)\phi(x^\prime)
=0\,,
\end{align}
that can be used to obtain the response of the system to a small
classical perturbation in the field $\phi(x)$.

\section{Quartic Coupling: The Seagull Diagram}
\label{section:one-loop}

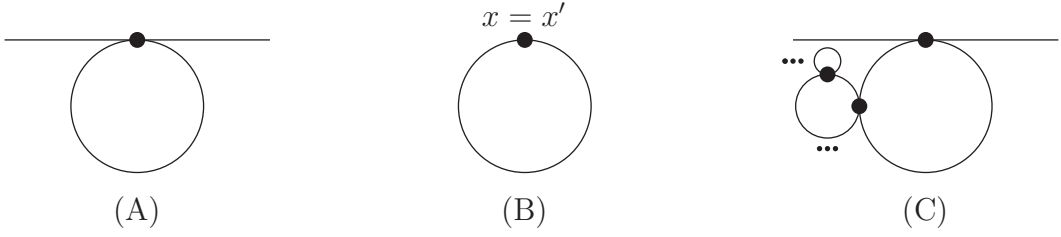
\begin{figure}[t!]
\begin{center}
\begin{picture}(100,75)(0,0)
\SetOffset(0,15)
\Line(0,50)(100,50)
\CArc(50,25)(25,0,360)
\Vertex(50,50){3}
\Text(50,-15)[]{(A)}
\end{picture}
\hskip 1.5cm
\begin{picture}(100,60)(0,0)
\SetOffset(0,15)
\CArc(50,25)(25,0,360)
\Vertex(50,50){3}
\Text(50,60)[]{$x=x^\prime$}
\Text(50,-15)[]{(B)}
\end{picture}
\hskip 1.5cm
\begin{picture}(100,75)(0,0)
\SetOffset(5,15)
\Line(0,50)(100,50)
\CArc(50,25)(25,0,360)
\Vertex(50,50){3}
\CArc(13,25)(12,0,360)
\Vertex(25,25){3}
\CArc(13,42)(5,0,360)
\Vertex(13,37){3}
\Vertex(-3,42){1}
\Vertex(0,42){1}
\Vertex(3,42){1}
\Vertex(-3,42){1}
\Vertex(10,9){1}
\Vertex(13,9){1}
\Vertex(16,9){1}
\Text(50,-15)[]{(C)}
\end{picture}
\end{center}
\caption{
\label{figure:seagull}
(A): Seagull diagram, (B): seagull contribution to the
self-energy, (C): daisies/super-daisies}
\end{figure}

In this Section, we consider the quartic self-interaction
\begin{align}
{\cal L}_{\rm int}=-\sqrt{-g}\frac{\lambda}{4!}\phi^4\,.
\end{align}
The seagull contribution to the self energy is~[{\it cf.}
Figure~\ref{figure:seagull}~(B), where the propagator ${\rm i}\Delta$
is understood to be dressed]
\begin{align}
\label{selferg:quartic}
\Pi^{fg}(x,x^\prime)=
\frac\lambda2
\sqrt{-g(\eta)}\delta^{fg}\delta^4(x-x^\prime){\rm i}\Delta^{fg}(x;x^\prime)\,.
\end{align}
This self energy is diagonal on the CTP, and therefore
\begin{subequations}
\begin{align}
\Pi^A&=\Pi^T-\Pi^>=\Pi^<-\Pi^{\bar T}\equiv\Pi^T\,,\\
\Pi^R&=\Pi^T-\Pi^<=\Pi^>-\Pi^{\bar T}\equiv\Pi^T\,.
\end{align}
\end{subequations}
The UV divergences of ${\rm i}\Delta(x;x^\prime)$ at the coincident point
is cancelled by appropriate counterterms in $\delta{\cal L}$.

Let us now make the \emph{ansatz} that the dressed propagator ${\rm i}\Delta^{fg}(x;x^\prime)$ takes the form of the propagator for a free field
with a mass $m_{\rm dyn}$ to be determined. This means that ${\rm i}\Delta^{fg}(x;x^\prime)$ satisfies
\begin{align}
\label{ansatz:Delta:mdyn}
a^4(-\nabla^2_x-m^2_{\rm dyn}){\rm i}\Delta^{fg}(x;x^\prime)
=fg \delta^{fg}{\rm i}\delta^4(x;x^\prime)\,,
\end{align}
implying in turn that $\Delta(x;x^\prime)$ takes the form of Eq.~(\ref{Delta0})
with $m^2$ replaced by $m^2_{\rm dyn}$. This ansatz therefore
accounts for the IR enhancement to its leading orders
$H^2/m^2_{\rm dyn}$ and $(H^2/m^2_{\rm dyn})^0$. Effectively, we assume that the growing term $\log(a(\eta)a(\eta'))$ in (\ref{Delta:m=0}) which leads to an IR divergence and a breakdown of perturbation theory when used in loop diagrams, can be resummed into a mass term.
Within the
present approximation, this means that we can use
\begin{align}
\label{Pi:quartic}
\Pi^T=\Pi^R
=\sqrt{-g}\lambda\frac{3 H^4}{16\pi^2 m_{\rm dyn}^2}\delta^4(x-x^\prime)
\,,
\end{align}
as the appropriate approximation
in order to extract the leading effects of the IR enhancements.
Since this self energy is manifestly local and moreover diagonal on the
CTP [{\it cf.} Eq.~(\ref{selferg:quartic})], the Schwinger-Dyson
equations take a very simple form, namely
\begin{align}
a^4(-\nabla^2_x-m^2){\rm i}\Delta^{ab}(x;x^\prime)
-\sqrt{-g}\lambda\frac{3H^4}{16\pi^2 m^2_{\rm dyn}}
{\rm i}\Delta^{ab}(x;x^\prime)
=\delta^{ab}{\rm i}\delta^4(x-x^\prime)\,.
\end{align}
Taking this point of view and using (\ref{ansatz:Delta:mdyn}), we obtain for $m^2_{\rm dyn}$
\begin{align}
\label{eq:selfcon}
m_{\rm dyn}^2=m^2+ \lambda\frac{3 H^4}{16\pi^2{m^2_{\rm dyn}}}\,,
\end{align}
such that
\begin{align}
m_{\rm dyn}^2=\frac{m^2}{2}+\sqrt{\frac{m^4}{4}+\lambda\frac{3H^4}{16\pi^2}}\,.
\end{align}
For $m\to0$, this reduces to
\begin{align}
\label{selfmass}
m_{\rm dyn}^2
=\frac{\sqrt{3\lambda}}{4\pi}H^2
\end{align}
in agreement with
Refs.~\cite{Starobinsky:1994bd,Riotto:2008mv,Burgess:2009bs}.
Note that the
requirement $m_{\rm dyn}\ll H$ is met provided $\lambda\ll 1$.

Turning our attention to the one-point function, we see that its effective equation of motion is consistent with the Schwinger-Dyson equation within our present
approximations. Substituting Eq.~(\ref{Pi:quartic}) into the effective equation
of motion~(\ref{EOM:effective}) for $\phi$, we obtain
\begin{align}
\label{EOM:effective:quartic}
a^4(-\nabla_x^2-m^2)\phi(x)
-\sqrt{-g}\lambda\frac{3 H^2}{16\pi^2 m_{\rm dyn}^2}
\phi(x)
=0\,.
\end{align}
This equation can be solved by using a plane wave ansatz
\begin{align}
\label{plane:wave}
\phi(x)=-\frac{\nabla^2_x}{m_{\rm dyn}^2}\phi(x)
+\phi(x)\times O\left(\frac{m^2_{\rm dyn}}{H^2}\right)
\end{align}
from which we can directly deduce again
Eqs.~(\ref{eq:selfcon},~\ref{selfmass}).

In closing this Section, we note that
the solution of the Schwinger-Dyson equation including the seagull
correction in terms of the dressed propagator
effectively resums the daisy and superdaisy corrections
in terms of the bare propagators,
as it is indicated in Figure~\ref{figure:seagull}~(C).
Moreover, the Schwinger-Dyson equations resum all one-particle
irreducible (1PI) contributions to the self-energies. Besides,
one should expect that certain additional higher-loop order
diagrams contribute to a complete self-consistent
determination of $m_{\rm dyn}$~\cite{Burgess:2010dd}.
When promoting $\phi(x)$
to a globally $O(N)$-symmetric field, these higher order
diagrams are generically suppressed by powers of
$1/N$~\cite{Starobinsky:1994bd,Riotto:2008mv,Burgess:2009bs,Berges:2001fi,Berges:2004yj}.
In Section~\ref{section:sunset}, we demonstrate however, that
even without resorting to $1/N$ expansion,
there is only one remaining diagram that contributes to
$m_{\rm dyn}$ at leading order in the IR-enhancement, which
is the sunset diagram.

\section{Triscalar Coupling}
\label{section:triscalar}

\begin{figure}[t!]
\begin{center}
\begin{picture}(100,60)(0,0)
\SetOffset(0,15)
\Line(0,25)(25,25)
\Line(75,25)(100,25)
\CArc(50,25)(25,0,360)
\Vertex(25,25){3}
\Vertex(75,25){3}
\Text(50,-15)[]{(A)}
\end{picture}
\hskip 1.5cm
\begin{picture}(100,60)(0,0)
\SetOffset(0,15)
\CArc(50,25)(25,0,360)
\Vertex(25,25){3}
\Vertex(75,25){3}
\Text(15,25)[]{$x$}
\Text(85,25)[]{$x^\prime$}
\Text(50,-15)[]{(B)}
\end{picture}
\end{center}
\caption{\label{figure:triscalar}
(A): Triscalar correction to the propagator, (B): triscalar
self energy.}
\end{figure}
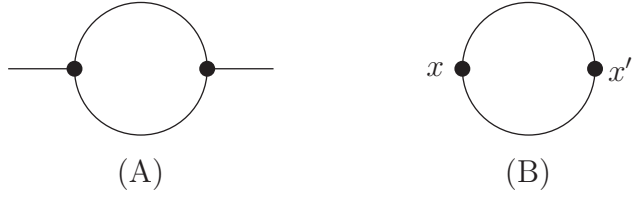

In order to distinguish better between the cause and the
effect of the dynamical mass term, we first introduce
an additional light minimally coupled scalar field $\chi$
of mass $m_\chi$, that couples to $\phi$ through the interaction
\begin{align}
\label{L:trisc:chi}
{\cal L}_{\rm int}=-\sqrt{-g}\frac{h_\chi}{2}\phi\chi^2\,.
\end{align}
Let us first look at the one-point function. The self energy of $\phi$ is [{\it cf.}
Figure~\ref{figure:triscalar}~(B)]
\begin{align}
\label{Pi:triscalar:CTP}
{\rm i}\Pi^{fg}(x;x^\prime)=
\frac 12 h_\chi^2\sqrt{-g(\eta)}\sqrt{-g(\eta^\prime)}
{\rm i}\Delta_\chi^{fg}(x;x^\prime){\rm i}\Delta_\chi^{fg}(x;x^\prime)\,.
\end{align}
The leading contribution in orders of $m_{\chi}^2/H^2$ to the
retarded self-energy is
\begin{align}
\label{Pi:triscalar}
\Pi^R(x;x^\prime)=
\frac 12 h_\chi^2\sqrt{-g(\eta)}\sqrt{-g(\eta^\prime)}
\frac{3H^4}{4\pi^2 m^2_\chi}\bar\Delta_\chi^{R}(x;x^\prime)
+O\left[\left(\frac{m_\chi^2}{H^2}\right)^0\right]\,.
\end{align}
Note that the retarded propagator $\bar\Delta_\chi^{R}$ is not identical
to $\Delta_\chi^{R}$, but it deviates as
\begin{align}
\label{Delta0:approx}
\bar\Delta_\chi^{R}=\Delta_\chi^{R}+ H^2
\times O\left(\frac{m^2_\chi}{H^2}\right)\,.
\end{align}
To the level of our approximations, {\it i.e.} the expansion
in $m^2_\chi/H^2$, we may use
\begin{align}
{\rm i}\bar\Delta_\chi^{R}=
\frac{H^2}{4\pi^2}
\left\{
-\frac{1}{y_{++}}+\frac{1}{y_{+-}}
-\frac12\log(y_{++})+\frac12\log(y_{+-})
\right\}\,.
\end{align}
The bar therefore indicates that it is accurate up to order
$(m_\chi^2/H^2)^0$.

Substituting this into Eq.~(\ref{EOM:effective}) and using
the methods employed {\it e.g.} in Ref.~\cite{Prokopec:2003tm}
does not lead to a local effective equation, however. We obtain
\begin{align}
\label{Eff:Eq:phi:cubed}
a^4(-\nabla_x^2-m^2)\phi(x)
-\frac12 h_\chi^2 \frac{3 H^4}{4\pi^2 m^2_\chi} a^4
\int d^4 x^\prime \sqrt{-g(x^\prime)}
\left[\bar\Delta^{R}(x;x^\prime)
+O(m_\chi^2)
\right]
\phi(x^\prime)=0\,.
\end{align}
The covariant d'Alembertian derivative acting on
$\bar\Delta_\chi^{R}$ is well defined,
\begin{align}
-a^4\nabla^2_x\bar\Delta_\chi^{R}(x;x^\prime)
=\delta^4(x-x^\prime)\,,
\end{align}
which implies that $\bar\Delta_\chi^{R}$ is the retarded
Green function for $-a^4\nabla^2_x$.
When substituting the plane wave ansatz~(\ref{plane:wave}),
we can perform the integral in Eq.~(\ref{Eff:Eq:phi:cubed}) by parts and obtain
\begin{align}
\label{Eff:Eq:phi:cubed:local}
a^4(-\nabla_x^2-m^2)\phi(x)&
-\frac12 h_\chi^2 \frac{3H^4}{4\pi^2 m^2_\chi}
\frac{1}{m^2_{\rm dyn}}\phi(x)
\\\notag
&+\frac12 h_\chi^2 \frac{H^4}{4\pi^2 m^2_\chi} a^4
\int d^4 x^\prime\frac{1}{m^2_{\rm dyn}}
\left[
\nabla^2_x O(m_\chi^2)
\right]\phi(x^\prime)=0\,.
\end{align}
The last term originates from the approximation~(\ref{Delta0:approx})
and it is subdominant compared to the second one by a factor of
$m^2_\chi/H^2$,
when assuming that
$m^2_\chi/H^2\ll1$ and $m^2_{\rm dyn}\sim m^2_\chi$.

The plane wave solution~(\ref{plane:wave}) therefore must satisfy
\begin{align}
\label{mdyn:triscalar}
m^2_{\rm dyn}
=m^2+\frac12 h_\chi^2\frac{3H^4}{4\pi^2 m^2_\chi}\frac{1}{m^2_{\rm dyn}}\,.
\end{align}
Replacing the interaction Lagrangian~(\ref{L:trisc:chi}) by the
cubic interaction
\begin{align}
{\cal L}_{\rm int}=-\sqrt{-g}\frac{h}{3!}\phi^3\,,
\end{align}
effectively leads to the replacements
$m_\chi^2\equiv m_{\rm dyn}^2$ and $h_\chi\to h$.
We then obtain an equation for a self-consistent mass:
\begin{align}
\label{selfcon:m:mdyn}
m_{\rm dyn}^6-m^2 m_{\rm dyn}^4-\frac 12 h^2 \frac{3H^4}{4\pi^2}=0.
\end{align}
When $m$ is negligibly small, this implies
\begin{align}
\label{selfcon:mdyn:only}
m^2_{\rm dyn}=\left(\frac{3}{8\pi^2}\right)^{1/3}
h^{2/3} H^{4/3}\,.
\end{align}
In order to meet the condition $m_{\rm dyn}\ll H$, we have to
require here that $h\ll H$.

An alternative way to derive these results is to act on
Eq.~(\ref{Eff:Eq:phi:cubed}) with $a^4\nabla^2_x a^{-4}$ from
the left. This approach is similar to the one pursued in
Ref.~\cite{Garbrecht:2006jm} for Yukawa interactions.
We obtain
\begin{align}
\label{eq:eff:phicubed}
a^4(-\nabla_x^2-m^2)\nabla_x^2\phi(x)
-a^4\frac 12 h^2 \frac{3 H^4}{4\pi^2 m^2_\chi}\phi(x)=0\,.
\end{align}
Substituting again the plane-wave ansatz~(\ref{plane:wave})
leads to Eq.~(\ref{Eff:Eq:phi:cubed:local}).

The above results for $m^2_{\rm dyn}$ can also be obtained by considering the Schwinger-Dyson equations for the propagator. However, in contrast to the one-loop self energy from the quartic
coupling, the self energy from the triscalar coupling is non-local and we therefore need to modify the procedure for solving the
Schwinger-Dyson equations. It is useful here to
split these into the Kadanoff-Baym equations and equations
for the retarded and advanced propagators as
\begin{subequations}
\begin{align}
a^4(-\nabla^2_x-m^2){\rm i}\Delta^{<,>}(x;x^\prime)=&
-{\rm i}\int d^4w\,{\rm i}\Pi^R(x;w){\rm i}\Delta^{<,>}(w,x^\prime)
\\\notag
&
-{\rm i}\int d^4w\,{\rm i}\Pi^{<,>}(x;w){\rm i}\Delta^A(w,x^\prime)
\,,
\label{kadanoff-Baym}\\
a^4(-\nabla^2_x-m^2){\rm i}\Delta^{R,A}(x;x^\prime)=&
{\rm i}\delta^4(x;x^\prime)
-{\rm i}\int d^4w\,{\rm i}\Pi^{R,A}(x;w){\rm i}\Delta^{R,A}(w,x^\prime)
\,.
\end{align}
\end{subequations}
Due to the $\delta$-function on the right-hand side, acting with
the d'Alembertian on the equations for the retarded and advanced
propagators does not lead to a useful simplification. We recall
from Section~\ref{section:CTP}
however that the retarded and advanced propagators do not contain
information about the mass at the leading orders $(m^2/H^2)^{-1}$
and $(m^2/H^2)^{0}$, such that we do not need these equations in
order to determine the self-consistent mass $m_{\rm dyn}$.
The Kadanoff-Baym equation (\ref{kadanoff-Baym}) can be reduced to
a local form when acting with the covariant d'Alembertian
$a^4\nabla_x^2a^{-4}$ from the left. Note that
$\nabla^2_x a^{-4}\Pi^{<,>}=h^2H^6\times O(1)$, such that this term
is negligible compared to the leading self-energy correction
$\nabla_x^2\int d^4 w\,\Pi^R(x,w)\sim h^2 H^4/m_{\rm dyn}^2$
when $m_{\rm dyn}\ll H$. In analogy with
Eq.~(\ref{eq:eff:phicubed}), we obtain
\begin{align}
\label{SD:triscalar:local}
a^4(-\nabla_x^2-m^2)
\nabla_x^2\Delta^{<,>}(x;x^\prime)
-a^4\frac12 h^2 \frac{3 H^4}{4\pi^2 m_{\rm dyn}^2}\Delta^{<,>}(x;x^\prime)=0\,.
\end{align}
Again, the plane wave
ansatz $(-\nabla_x^2-m^2_{\rm dyn})\Delta^{<,>}(x;x^\prime)=0$
then leads to the self-consistency
relations~(\ref{selfcon:m:mdyn},~\ref{selfcon:mdyn:only})
for $m_{\rm dyn}$.

We emphasise that this local form holds only for the
plane-wave ansatz
with time-independent $m^2_{\rm dyn}$. In general, we expect that
during the time when the IR terms are accumulating as $\log(a(x)a(x^\prime))$, the ansatz of a time-independent $m_{\rm dyn}$ should not be
valid and the memory integrals cannot be simplified to an effectively local
form.

In order to compare with the earlier works on the
self-energies of photons~\cite{Prokopec:2003tm} and of
chiral fermions~\cite{Garbrecht:2006jm}, which are conformally
coupled particles, we conclude this Section by considering
the situation when $\phi$ is conformally coupled to the
expanding background, which can be achieved by adding the
term $-\frac16\sqrt{-g}R\phi^2$ to the
Lagrangian~(\ref{Lagrangian}), where $R$ is the Ricci scalar,
which takes the value $R=12H^2$ in de~Sitter space.
We again couple to a minimally coupled light ($m_\chi\ll H$)
field $\chi$
through the interaction~(\ref{L:trisc:chi}).

The effective equation of motion then reads
\begin{align}
a^4(-\nabla^2_x-m^2-\frac16 R)\phi(x)
-\int d^4x^\prime\,\Pi^R(x;x^\prime)\phi(x)=0\,,
\end{align}
with $\Pi$ as in Eqs.~(\ref{Pi:triscalar:CTP},~\ref{Pi:triscalar}).
Acting with $a^4\nabla_x^2a^{-4}$ from the left, we obtain
\begin{align}
\label{EOM:conformal:local}
a^4(-\nabla^2_x-m^2-\frac16 R)\nabla_x^2\phi(x)
-a^4\frac12 h_\chi^2 \frac{3H^4}{4\pi^2m_\chi^2}\phi(x)=0\,.
\end{align}
The plane-wave ansatz for a conformally coupled, massive field
\begin{align}
\left[-\nabla_x^2-\frac16 R\right]\phi(x)=m^2_{\rm dyn}\phi(x)
\end{align}
leads to
\begin{align}
(m^2_{\rm dyn}-m^2)(m_{\rm dyn}^2+\frac16 R)
=\frac 12 h_\chi^2 \frac{3H^4}{4\pi^2 m_\chi^2}\,.
\end{align}
For $m^2\ll m_{dyn}^2$ and $m_{\rm dyn}^2\ll H^2$, we find
\begin{align}
\label{mass:conformal}
m_{\rm dyn}^2=h_\chi^2\frac{3 H^2}{16\pi^2m_\chi^2}\,.
\end{align}
Compared to the mass-square for the minimally coupled scalar
field~(\ref{mdyn:triscalar}) with $m=0$,
the right hand
side differs by a factor of $m_{\rm dyn}^2/H^2$. It is therefore
interesting to compare this result with earlier ones obtained
for conformally coupled fields interacting with a light, minimally coupled
scalar. For a photon $\gamma$ coupled to
a scalar field with charge $e$ and mass $m_\chi$, the mass-square is
\begin{align}
\label{mass:photon}
m_\gamma^2=\frac{3e^2H^4}{4\pi^2 m_\chi^2}\,,
\end{align}
while for a pair of Weyl fermions $\psi_{L,R}$ with a Yukawa coupling
$f$ to a real scalar field of mass $m_\chi$
\begin{align}
\label{mass:fermion}
m^2_\psi=\frac{3 f^2 H^4}{8\pi^2 m_\chi^2}
\end{align}
is found for the dynamical
effective mass-square. Up to a replacement of a factor
$H^2\to h_\chi^2$, that takes place for dimensional reasons, the
result~(\ref{mass:conformal})
for a conformally coupled scalar field
fits well into one scheme
with the earlier results for photons~(\ref{mass:photon}) and
fermions~(\ref{mass:fermion}), which are both conformally coupled
as well, while Eq.~(\ref{mdyn:triscalar})
for the mass square of the minimally coupled scalar field
exhibits an interesting difference.

\section{Quartic Coupling: The Sunset Diagram}
\label{section:sunset}

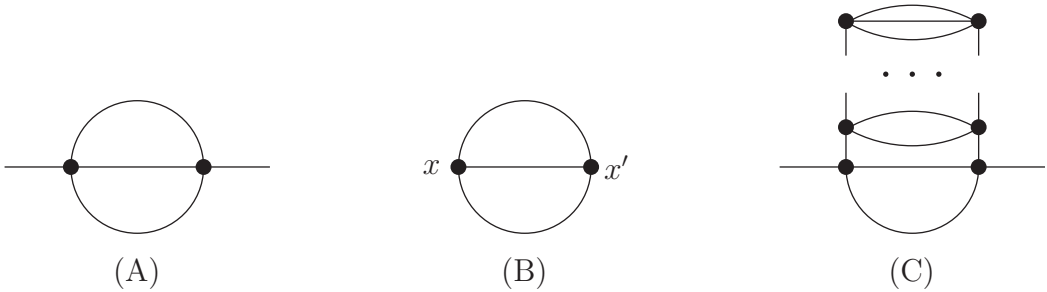
\begin{figure}[t!]
\begin{center}
\begin{picture}(100,60)(0,0)
\SetOffset(0,15)
\Line(0,25)(100,25)
\CArc(50,25)(25,0,360)
\Vertex(25,25){3}
\Vertex(75,25){3}
\Text(50,-15)[]{(A)}
\end{picture}
\hskip 1.5cm
\begin{picture}(100,60)(0,0)
\SetOffset(0,15)
\Line(25,25)(75,25)
\CArc(50,25)(25,0,360)
\Vertex(25,25){3}
\Vertex(75,25){3}
\Text(15,25)[]{$x$}
\Text(85,25)[]{$x^\prime$}
\Text(50,-15)[]{(B)}
\end{picture}
\hskip 1.5cm
\begin{picture}(100,90)(0,0)
\SetOffset(0,15)
\Line(0,25)(100,25)
\CArc(50,25)(25,180,360)
\Vertex(25,25){3}
\Vertex(75,25){3}
\Line(25,25)(25,53)
\Line(75,25)(75,53)
\CArc(50,83)(50,240,300)
\CArc(50,-4)(50,60,120)
\Vertex(25,40){3}
\Vertex(75,40){3}
\Vertex(40,60){1}
\Vertex(50,60){1}
\Vertex(60,60){1}
\Line(25,67)(25,80)
\Line(75,67)(75,80)
\Line(25,80)(75,80)
\CArc(50,123)(50,240,300)
\CArc(50,36)(50,60,120)
\Vertex(25,80){3}
\Vertex(75,80){3}
\Text(50,-15)[]{(C)}
\end{picture}
\end{center}
\caption{\label{figure:sunset}
(A): Sunset diagram, (B): sunset contribution to the
self-energy, (C): ladder diagram.}
\end{figure}

The methods for finding the effective
self-consistent mass for the cubic interaction
can also be applied to evaluate the leading contribution to the sunset
diagram of $\lambda \phi^4$ in terms of an effective mass term.
This is because the leading contribution to the retarded
self-energy contains the IR-enhancement term to quadratic order
and again, the retarded Green function of the covariant
d'Alembertian as additional factor.
Explicitly,
the leading ($\sim 1/m_{\rm dyn}^4$) contribution to the self-energy
is~[{\it cf.} Figure~\ref{figure:sunset}~(B)]
\begin{align}
{\rm i}\Pi^R(x;x^\prime)=
\frac{\lambda^2}{2}
\left(
\frac{3 H^2}{8\pi^2 m_{\rm dyn}^2}
\right)^2
{\rm i}\bar\Delta^R(x;x^\prime)
\end{align}
and therefore, using the plane wave ansatz~(\ref{plane:wave})
\begin{align}
\label{PiR:sunset}
\int d^4 x^\prime
\Pi^R(x;x^\prime)\phi(x^\prime)
=\frac{\lambda^2}{2}a^4\left(\frac{3 H^4}{8\pi^2 m_{\rm dyn}^2}\right)
\frac{1}{m^2_{\rm dyn}}\phi(x)\,.
\end{align}
This contribution is to be combined with the one-loop result
from Eq.~(\ref{EOM:effective:quartic}), what leads to the self-consistency
relation
\begin{align}
\frac{\lambda^2}{2}\left(\frac{3H^4}{8\pi^2 m^2_{\rm dyn}}\right)^2
\frac{1}{m^2_{\rm dyn}}
+\lambda\frac{3 H^4}{16\pi^2 m^2_{\rm dyn}}=a^4 m_{\rm dyn}^2\,.
\end{align}
We find
\begin{align}
\label{selfcon:one-loop:sunset}
m^2_{\rm dyn}=\frac{\sqrt{3\lambda}H^2}{2\sqrt2\pi}\,.
\end{align}
Note that both, the one-loop and the sunset diagram contribute at
the same order in $\lambda$.
This relation can also be obtained from Schwinger-Dyson equations,
in complete analogy with Section~\ref{section:triscalar}, such that
we do not repeat this discussion here.

We note that the iterative addition of two rungs to the sunset diagram
leads to additional ladder diagrams as indicated
in Figure~\ref{figure:sunset}~(C).
Each such insertion is suppressed by
a factor of $\lambda^2$ from the new vertices, but it is also enhanced
by a factor of $\lambda^2$ from the IR-enhanced contributions of the four
new propagators, which are $\sim 1/\sqrt{\lambda}$ each. However,
these ladder diagrams are not part of the proper self-energy. This is most
easily seen from the definition~(\ref{def:Pi:proper}).
The diagram contributing to $\Gamma_2$, which gives rise to
the sunset contribution to $\Pi$, is 2PI, whereas the
vacuum ladder diagrams are two-particle reducible and therefore not part
of $\Gamma_2$.
We therefore conclude that the ladder
diagrams are readily resummed within the self-consistency
relation~(\ref{selfcon:one-loop:sunset}) (along with super ladders
arising from ladder insertions into internal lines).

\begin{figure}[t!]
\begin{center}
\begin{picture}(100,60)(0,0)
\SetOffset(0,15)
\Line(0,25)(25,25)
\Line(75,25)(100,25)
\CArc(50,25)(25,0,360)
\CArc(25,0)(25,0,90)
\CArc(75,0)(25,90,180)
\Vertex(25,25){3}
\Vertex(75,25){3}
\Vertex(50,0){3}
\Text(50,-15)[]{(A)}
\end{picture}
\hskip 1.5cm
\begin{picture}(100,60)(0,0)
\SetOffset(0,15)
\Line(0,25)(100,25)
\CArc(50,25)(25,0,360)
\CArc(29,37.5)(25,327,33)
\CArc(71,37.5)(25,147,213)
\Vertex(25,25){3}
\Vertex(75,25){3}
\Vertex(50,50){3}
\Vertex(50,25){3}
\Text(50,-15)[]{(B)}
\end{picture}
\end{center}
\caption{
\label{figure:NLO:NNLO}
Diagram~(A) contributes at NLO in the $1/N$ expansion,
diagram~(B) at NNLO. However, the IR-enhanced contributions
from these diagrams are suppressed in the expansion in
$m^2_{\rm dyn}/H^2$.}
\end{figure}
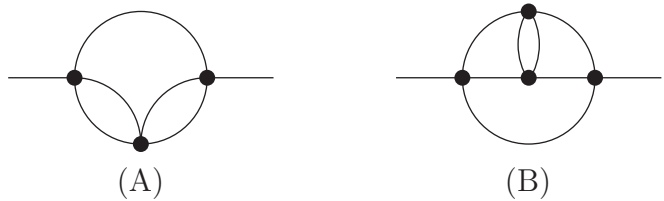

By this counting argument, according to which each
vertex contributes a factor $\lambda$ and each propagator
a factor $1/\sqrt\lambda$, the diagrams in
Figure~\ref{figure:NLO:NNLO} might be expected to contribute
to the self-consistency relation that determines
$m_{\rm dyn}$  at the leading (negative) order in
$m^2_{\rm dyn}/H^2$ as well, {\it cf. e.g.} Ref.~\cite{Burgess:2010dd}.
When the value of the propagator is non-perturbatively large,
one can consider a theory with an $O(N)$ symmetric field $\phi(x)$
and perform a $1/N$ expansion~\cite{Berges:2001fi,Berges:2004yj}.
In this situation, the seagull graph
of Figure~\ref{figure:seagull}~(A) is leading order
[LO, $O(1)$].
The sunset diagram and an infinite series
of diagrams that starts with the one in
Figure~\ref{figure:NLO:NNLO}~(A) contributes
at next to leading order [NLO, $O(1/N)$].
The diagram of Figure~\ref{figure:NLO:NNLO}~(B)
is one of infinitely many diagrams that contribute
at the next-to next-to leading order [NNLO, $O(1/N^2)$].
While it is possible to sum all NLO diagrams in such a theory,
we notice that this is not necessary in the present case,
because the diagrams in
Figure~\ref{figure:NLO:NNLO} do not contribute
to the retarded self-energy at leading (negative)
order in $m^2_{\rm dyn}/H^2$: At the leading order in
$m^2_{\rm dyn}/H^2$, all except for one propagator are IR enhanced.
This is because when all propagators are IR enhanced, the summation
over CTP indices $\pm$
(taking account of a factor of $\pm 1$ for these vertices
and of the fact that the IR enhanced term is identical for all
four CTP propagators)
immediately leads to a vanishing result.
By the same token, if the remaining non-IR-enhanced propagator
is not directly connected to the right external vertex of the
self-energy, the CTP summation again leads to a vanishing
contribution to the retarded self-energy.
Now, if the single non-IR-enhanced propagator
is connected to an internal vertex and
the right external vertex, the identity
$\Delta^{++}+\Delta^{--}-\Delta^{+-}-\Delta^{-+}=0$
leads to a vanishing contribution to the retarded self-energy.
Finally, we consider the case where the non-IR-enhanced propagator
connects to both external vertices. If there are internal vertices,
the CTP sum leads to a vanishing term again. Only in the case
of the sunset diagram, where there are no internal vertices,
a non-vanishing contribution remains.

\section{Discussion and Conclusions}
\label{section:conclusions}

In this work, we have calculated retarded self-energies
for scalar fields with quartic and trilinear interactions
at leading order in the IR enhancement. One of the main results
is Eq.~(\ref{selfcon:one-loop:sunset}), which can be interpreted
as the effective self-consistent mass that a scalar field
with quartic self-interaction acquires in a steady-state
de~Sitter space. Self consistent masses during inflation
have been determined earlier in Refs.~\cite{Starobinsky:1994bd,Riotto:2008mv,Burgess:2009bs}. In the present
paper, we have used the propagator~(\ref{Delta0}), which
has been proposed in
Ref.~\cite{Prokopec:2003tm} and which
perhaps leads to simplifications when compared to earlier
treatments. Moreover, we have determined the
effective self-mass for self-energies, that are non-local,
{\it i.e.} for the triscalar interaction and for the sunset
diagram. For this purpose, we have made a plane-wave ansatz
in a steady-state de~Sitter
background~(\ref{ansatz:Delta:mdyn},\ref{plane:wave}), which allows
to reduce the effective equations of motion for the propagator
and for the one-point function to a local form,
{\it cf.} Eqs.~(\ref{Eff:Eq:phi:cubed:local},\ref{eq:eff:phicubed},\ref{SD:triscalar:local},\ref{EOM:conformal:local}). In turn,
this implies that provided the system is not in a steady
state [substantial deviations from de~Sitter and/or deviations
from the plane wave
form~(\ref{ansatz:Delta:mdyn},\ref{plane:wave})
for the scalar field], there
remain in principle still non-local terms in the form of memory
integrals. In contrast, the self-energy from the seagull
diagram~(\ref{selferg:quartic}) is always manifestly local.
This perhaps explains differences between the cases of
a quartic and a trilinear self-interaction, that have
been observed in Ref.~\cite{Burgess:2009bs}
when using functional renormalisation
group methods.

The approximation strategies employed in the present work appear
related to those of Ref.~\cite{Rajaraman:2010xd} for a scalar field on a four-dimensional sphere. Because the sphere is compact, there
is a discrete zero mode. The leading self-consistent
propagator is obtained when including the loop effects
of the zero mode (including its own self-interactions), while
neglecting the loop effects of the remaining modes. Here, we extract
the leading contributions from the IR enhanced term ${3H^4}/{8\pi^2m^2}$ (which may be
mapped to the zero mode on the sphere) within the loop diagrams.
While both approaches appear to be very similar, there is still
a difference in the numerical coefficient of $m_{\rm dyn}$.
It would be interesting to resolve the origin of this discrepancy.

We briefly estimate whether the self-regulatory mass is sufficient to
bar large IR effects on $(\lambda/4!)\phi^4$ inflation.
From Eq.~(\ref{selfcon:one-loop:sunset}), we immediately see
that the dynamical mass is too small to change the phenomenology
of this model of inflation, because $\lambda\sim10^{-14}$ and
consequently $m_{\rm dyn}\ll H$ by orders of magnitude.
Using Eqs.~(\ref{Delta0}) and~(\ref{selfcon:one-loop:sunset}),
we can estimate the IR enhanced term [the third term
on the right hand side of Eq.~(\ref{Delta0})] as
\begin{align}
\label{phi2:IR}
\langle\phi^2\rangle_{\rm IR}\approx
\sqrt\frac32 \frac{H^2}{2\sqrt\lambda}\,.
\end{align}
Since $H=\sqrt{8\pi/3}\sqrt{(\lambda/4!)\phi^4/m_{\rm Pl}^2}$
and $\langle \phi \rangle =O(m_{\rm Pl})$ during inflation,
the IR enhanced term (\ref{phi2:IR}) is $\sim\sqrt\lambda m_{\rm Pl}^2$, which
is much smaller than $\langle \phi\rangle^2$. Finally, it is
interesting to notice that if we suppose $\langle\phi^2\rangle_{\rm IR}\approx 0$ initially and that it subsequently
grows as $[H^2/(8\pi^2)]N_{\rm e}$
(as may be motivated from
Refs.~\cite{Allen:1985ux,Allen:1987tz,Tsamis:1993ub})
it takes
$N_{\rm e}\approx2\sqrt6\pi/\sqrt\lambda$ e-folds
before saturating to the
value~(\ref{phi2:IR}).
We emphasize that these are rough estimates in particular
due to the fact that the self-mass~(\ref{selfcon:one-loop:sunset})
only applies to a steady state, while here we have discussed
properties of an evolving system. Sizeable effects from
the IR enhancement may yet be expected in some multi-field
models of inflation or small field models with a large
number of e-folds.

In conclusion, the present work
shows that there is a self-regulatory dynamical mass for
$\phi^4$ theory in de~Sitter which suppresses the IR enhanced
terms. The only leading IR enhanced contributions to
the retarded self-energy originate from the seagull and the
sunset diagrams and can be calculated in a self-consistent
way. It would be interesting to apply the techniques
presented here to study some aspects of gravitational interactions
in de~Sitter space as well as cosmological perturbations in slow-roll inflation.

\subsubsection*{Acknowledgements}

\noindent
The authors would like to thank Martin~Beneke, Emanuela~Dimastrogiovanni and David~Seery for useful discussions.
This work is supported by the Gottfried Wilhelm Leibniz programme
of the Deutsche Forschungsgemeinschaft.


\begin{thebibliography}{99}

\bibitem{Allen:1985ux}
  B.~Allen,
  ``Vacuum States in de Sitter Space,''
  Phys.\ Rev.\  {\bf D32 } (1985)  3136.


\bibitem{Allen:1987tz}
  B.~Allen, A.~Folacci,
  ``The Massless Minimally Coupled Scalar Field In De Sitter Space,''
  Phys.\ Rev.\  {\bf D35 } (1987)  3771.

\bibitem{Tsamis:1993ub}
  N.~C.~Tsamis, R.~P.~Woodard,
  ``The Physical basis for infrared divergences in inflationary quantum gravity,''
  Class.\ Quant.\ Grav.\  {\bf 11}, 2969-2990 (1994).

\bibitem{Seery:2010kh}
  D.~Seery,
  ``Infrared effects in inflationary correlation functions,''
  Class.\ Quant.\ Grav.\  {\bf 27 } (2010)  124005.
  [arXiv:1005.1649 [astro-ph.CO]].


\bibitem{Starobinsky:1994bd}
  A.~A.~Starobinsky, J.~Yokoyama,
  ``Equilibrium state of a selfinteracting scalar field in the De Sitter background,''
  Phys.\ Rev.\  {\bf D50 } (1994)  6357-6368.
  [astro-ph/9407016].

\bibitem{Riotto:2008mv}
  A.~Riotto, M.~S.~Sloth,
  ``On Resumming Inflationary Perturbations beyond One-loop,''
  JCAP {\bf 0804 } (2008)  030.
  [arXiv:0801.1845 [hep-ph]].

\bibitem{Burgess:2009bs}
  C.~P.~Burgess, L.~Leblond, R.~Holman, S.~Shandera,
  ``Super-Hubble de Sitter Fluctuations and the Dynamical RG,''
  JCAP {\bf 1003 } (2010)  033.
  [arXiv:0912.1608 [hep-th]].

\bibitem{Rajaraman:2010xd}
  A.~Rajaraman,
  ``On the proper treatment of massless fields in Euclidean de Sitter space,''
  Phys.\ Rev.\  {\bf D82 } (2010)  123522.
  [arXiv:1008.1271 [hep-th]].


\bibitem{Prokopec:2003tm}
  T.~Prokopec, E.~Puchwein,
  ``Photon mass generation during inflation: de Sitter invariant case,''
  JCAP {\bf 0404 } (2004)  007.
  [astro-ph/0312274].


\bibitem{Garbrecht:2006jm}
  B.~Garbrecht, T.~Prokopec,
  ``Fermion mass generation in de Sitter space,''
  Phys.\ Rev.\  {\bf D73 } (2006)  064036.
  [gr-qc/0602011].

\bibitem{Onemli:2002hr}
  V.~K.~Onemli, R.~P.~Woodard,
  ``Superacceleration from massless, minimally coupled $\phi^4$,''
  Class.\ Quant.\ Grav.\  {\bf 19 } (2002)  4607.
  [gr-qc/0204065].

\bibitem{Onemli:2004mb}
  V.~K.~Onemli, R.~P.~Woodard,
  ``Quantum effects can render $w < -1$ on cosmological scales,''
  Phys.\ Rev.\  {\bf D70 } (2004)  107301.
  [gr-qc/0406098].

\bibitem{Brunier:2004sb}
  T.~Brunier, V.~K.~Onemli, R.~P.~Woodard,
  ``Two loop scalar self-mass during inflation,''
  Class.\ Quant.\ Grav.\  {\bf 22 } (2005)  59-84.
  [gr-qc/0408080].

\bibitem{Garbrecht:2006df}
  B.~Garbrecht,
  ``Ultraviolet Regularisation in de Sitter Space,''
  Phys.\ Rev.\  {\bf D74 } (2006)  043507.
  [hep-th/0604166].

\bibitem{Burgess:2010dd}
  C.~P.~Burgess, R.~Holman, L.~Leblond, S.~Shandera,
  ``Breakdown of Semiclassical Methods in de Sitter Space,''
  JCAP {\bf 1010 } (2010)  017.
  [arXiv:1005.3551 [hep-th]].

\bibitem{Berges:2001fi}
  J.~Berges,
  ``Controlled nonperturbative dynamics of quantum fields out-of-equilibrium,''
  Nucl.\ Phys.\  {\bf A699 } (2002)  847-886.
  [hep-ph/0105311].

\bibitem{Berges:2004yj}
  J.~Berges,
  ``Introduction to nonequilibrium quantum field theory,''
  AIP Conf.\ Proc.\  {\bf 739 } (2005)  3-62.
  [hep-ph/0409233].


\end{thebibliography}
\end{document}